# Hydrogen sulfide at high pressure: change in stoichiometry

Alexander F. Goncharov,[1,2,3] Sergey Lobanov,[2,4] Ivan Kruglov,[5] Xiao-Miao Zhao,[2,6,7] Xiao-Jia Chen,[1,2,6] Artem R. Oganov,[5,8,9,10] Zuzana Konôpková,[11] Vitali Prakapenka[12]

[1]*Key Laboratory of Materials Physics, Institute of Solid State Physics, Chinese Academy of Sciences, Hefei, Anhui 230031, China*

[2]*Geophysical Laboratory, Carnegie Institution of Washington, Washington, D.C. 20015, USA*

[3]*University of Science and Technology of China, Hefei, 230026, China*

[4]*V.S. Sobolev Institute of Geology and Mineralogy, SB RAS, 3 Pr. Ac. Koptyga, Novosibirsk 630090, Russia*

[5]*Moscow Institute of Physics and Technology, Dolgoprudny, Moscow Region 141700, Russian Federation*

[6]*Center for High Pressure Science and Technology Advanced Research, Shanghai 201203, China,*

[7]*Department of Physics, South China University of Technology, Guangzhou 510640, China*

[8]*Skolkovo Institute of Science and Technology, Skolkovo Innovation Center, Moscow 143026, Russia*

[9]*Department of Geosciences, Center for Materials by Design, and Institute for Advanced Computational Science, State University of New York, Stony Brook, NY 11794-2100.*

[10]*Northwestern Polytechnical University, Xi'an 710072, China.*

[11]*DESY Photon Science, Notkestrasse 85, D-22607 Hamburg, Germany*

[12]*Center for Advanced Radiation Sources, University of Chicago, Chicago, Illinois 60637, USA*

**Hydrogen-sulfide ($H_2S$) was studied by x-ray synchrotron diffraction (XRD) and Raman spectroscopy up to 150 GPa at 180-295 K and by quantum-mechanical variable-composition evolutionary simulations. The experiments show that $H_2S$ becomes unstable with respect to formation of new compounds with different structure and composition, including *Cccm* and a body-centered-cubic (bcc) like (*R3m* or *Im-3m*) $H_3S$, the latter one predicted previously to show a record-high superconducting transition temperature, $T_c$ of 203 K. These experiments provide experimental ground for understanding of this record high $T_c$. The experimental results are supported by theoretical structure searches that suggest the stability of new $H_3S$, $H_4S_3$, $H_5S_8$, $H_3S_5$, and $HS_2$ compounds at elevated pressures.**

## I. Introduction

Hydrogen bearing materials hold a promise to possess very high temperature superconductivity due to high vibrational frequencies, strong electron-phonon interaction, and partially covalent bonding [1]. Novel polyhydrides predicted theoretically to become stable at high pressure [2, 3] are good candidates to realize such expectations but several experimental efforts were inconclusive [4-6]. Recently, however, following the theoretical prediction [3], superconductivity up to 203 K in the H-S system at above 144 GPa has been reported [7, 8]. In the absence of experimental data on the structure and composition of the superconducting phase, the insights were offered by several theoretical calculations [3, 9-14] suggesting that it is a new polyhydride $H_3S$ which possesses such unusually high $T_c$ due to the strong electron-phonon coupling and high hydrogen (H) phonon frequencies. This called for more detailed experimental investigation of high-pressure behavior of $H_2S$ as the existing structural and experimental data are scarce and partially inconsistent [15-18] concerning the structure, physical properties, and possible chemical reactivity of this system at high pressure.

The behavior of $H_2S$ is complex at high pressure (Fig. 1), which includes orientational molecular ordering, formation of dynamically disordered states, and finally transformation to atomic (nonmolecular) extended structures. Recently, however, it has been theoretically predicted that other than $H_2S$ compounds become stable at high pressures [9, 12, 19], the most stable of which is $H_3S$ [3, 9, 12, 19] that evolves under pressure from a mixed molecular-extended *Cccm* [3] structure to quasi-molecular (*R3m*) and finally to atomic (*Im-3m*) solid. The latter two phases are found to have a very high $T_c$ (160 and 190 K, respectively) and, if confirmed experimentally, their formation at high pressures may explain the observations of superconductivity above 100 GPa (up to 203 K) in experiments [7, 8].

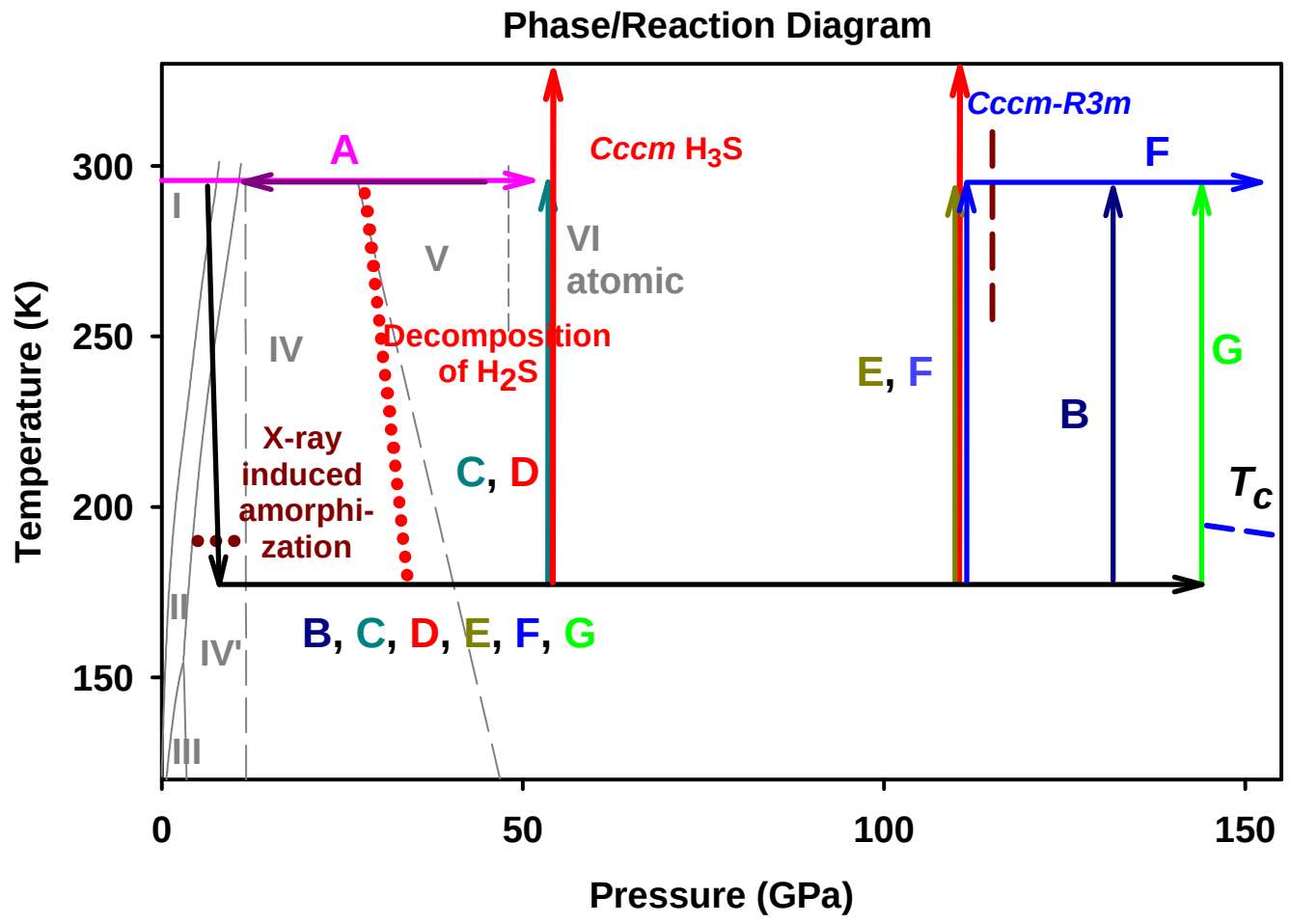

Fig. 1 (color online). Phase/reaction diagram of $H_2S$ superimposed on the experimental P-T paths (arrows of different colors) explored in this work. The physical phase boundaries shown by gray thin lines are from Ref. [17]. The pressure dependent $T_c$ is from Ref. [8]. The dotted lines, which correspond to the onset of chemical decomposition and low-temperature amorphization as well as *Cccm-R3m* boundary (dashed line) are from this work. The experiments D and E used laser annealing at 180 K (<1000 K) shown by red vertical arrows (not in scale).

Here, we report the results of low- (180 K) and room-temperature (295 K) X-ray diffraction (XRD) and Raman spectroscopy investigations to 150 GPa (Fig. 1) that reveal an increasing instability of $H_2S$ with pressure and formation of new $H_xS$ compounds at elevated pressures. At above 55 GPa we find *Cccm* $H_3S$ [3] and other unknown sulfur-rich crystalline material, while *R3m* (*Im-3m*) $H_3S$ has been recorded above 110 GPa. However, we found that the formation of this phase is impeded, which explain the observed previously dependence of $T_c$ on the experimental P-T path [7, 8]. The experiments have been combined with theoretical variable composition searches using USPEX code [20] that revealed new stable $H_3S$, $H_4S_3$, $H_5S_8$, $HS_2$, and

$H_3S_5$ compounds at 25-150 GPa.

## II. Experimental and computational methods

$H_2S$ was loaded cryogenically in diamond anvil cells (DACs) with rhenium gaskets. Diamonds anvils with a flat 300 and beveled 300/50-100 μm culet diameter were used in Raman spectroscopy and XRD experiments measurements at room temperature and low (as low as 180 K) temperatures. In totally, 7 experiments have been performed which differ in the final pressure, P-T path, and probe technique (Fig. 1 and Table 1). The main emphasis in our experiments was to reveal the structural and compositional changes in $H_2S$ at the 180 K compression that may be related to observation of high $T_c$ [8]. The low-temperature Raman and XRD runs were performed separately, but the samples warmed to 295 K were subsequently cross examined. The experiment A (at 295 K) used Raman probe on compression and XRD – on decompression, while experiments B through G used Raman (C, F, G) and XRD (B, D, E) probes separately at low temperatures and usually both probes - at 295 K. The experiments D and E used laser annealing at 180 K (<1000 K). The room-temperature XRD was studied on unloading from 51 GPa following the loading run where Raman and optical spectra were studied. The low-temperature XRD was studied using a cryo-stream $N_2$ refrigerator, while an optical cryostat with a cold finger was used in Raman experiments. The samples were laser heated in an extended area with a low power (<2 W from each side) while at low temperatures at 54 and 120 GPa in two separate pressure runs; the annealing at room temperature has been also performed with the sample warmed up after the low temperature XRD and Raman study. Pressure was determined from the ruby fluorescence and gold XRD pressure markers with the appropriate temperature corrections, and Raman of the stressed diamond. For the micro-Raman experiments, a backscattering geometry was adopted for confocal measurements with incident laser wavelengths of 488, 532, and 660 nm. The XRD experiments were collected at the synchrotron beam line, sector 13 of the Advanced Phonon Source (APS) of the Argonne National Laboratory and Extreme Conditions Beamline P02.2 at DESY (Germany) with the wavelengths of 0.310 and 0.3344 (GSECARS) and 0.2906 Å (DESY).

Predictions of stable phases were done using the USPEX code in the variable-composition mode [20] at 0, 30, 60, 90, 120, 140 GPa. The first generation of structures (up to 32 atoms per the primitive cell) was produced randomly and the succeeding generations were obtained by applying heredity, atom transmutation, softmutation, and random variational operators, with probabilities of 40%, 20%, 20%, and 20%, respectively. 70% non-identical structures of each generation with the lowest enthalpies were used to produce the next generation. All structures were relaxed using density functional theory (DFT) calculations within the Perdew-Burke-Ernzerhof (PBE) [21], as implemented in the VASP code [22]. The plane-wave energy cutoff was chosen as 320 eV and Γ-centered uniform k meshes with resolution $2\pi \times 0.05$ $A^{\circ\,-1}$ were used. Zero point motion was not included. After using VASP code all stable structures were relaxed at experimental pressures.

## III. Results

The room temperature experiment (A) showed changes in RS, which are consistent with previously detected phase changes [23] to orientionally ordered phase IV, and then to phases V and VI, where the molecular character is diminished [16, 18] (Fig. 2). Above 28 GPa, RS detected modes that correspond to elemental S and molecular $H_2$ at 4150 $cm^{-1}$; however above 51 GPa a broad Raman peak at 400 $cm^{-1}$ appeared, while signs of S and $H_2$ diminish. XRD of the sample A measured at 51 GPa and 295 K (Fig. 3(a)) showed a pattern which agrees fairly well with that reported previously [24] and was assigned to phase V. We find that the experimental XRD pattern matches reasonably well the *Pc* $H_2S$ structure proposed theoretically [19, 25] except there are a several peaks (*e.g.*, at $6.3^0$) that suggest the beginning of the chemical transformation and can be connected to appearance of the *P-1* $H_4S_3$ predicted in this work (Table 2, Fig. 3(a)). On the pressure release, we found that there is a back transformation to phase IV (confirmed also by RS), while all additional high-pressure Bragg reflections disappear. XRD of phase IV agrees well with the tetragonal *I41/acd* $H_2S$ (phase IV) proposed by Fujihisa et al. [23] (Fig. 3(b)).

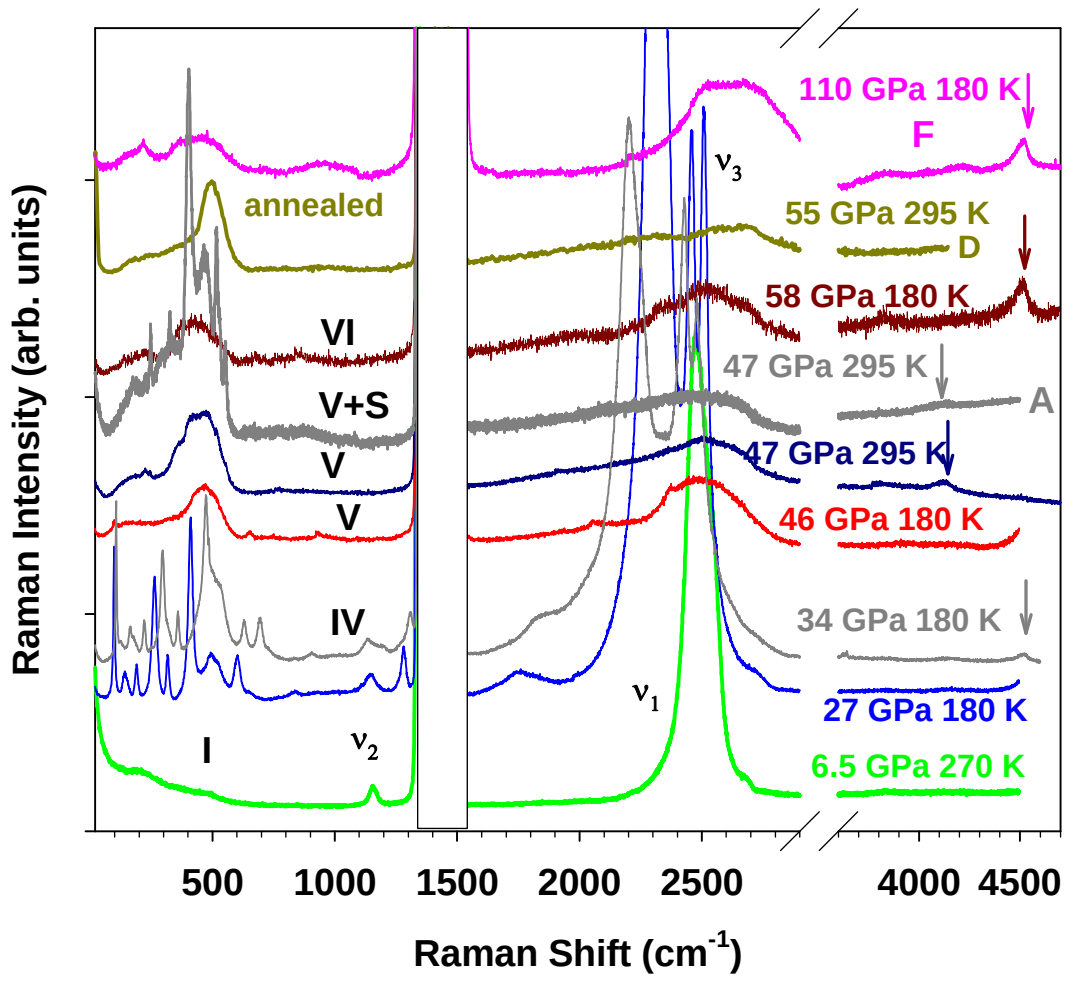

Fig. 2. Raman spectra of $H_2S$ on compression to 110 GPa. All curves are from experiment C except those which marked otherwise. Elemental S is detected in the spectrum at 47 GPa, 295 K via observations of narrow bands below 550 $cm^{-1}$; in this experiment (A) pressure was increased at 295 K unlike the other experiment (C), which reached the same P-T conditions via the 180 K path and subsequent warming up. Arrows indicate the $H_2$ vibron modes that appear under pressure as the result of $H_2S$ molecular decomposition. The sample at 55 GPa (D) is measured after a gentle laser heating which results in formation of *Cccm* $H_3S$.The first order Raman of diamond is masked by a rectangular.

When compressing at 180 K, Raman experiments (C, F, and G) showed a similar phase change as at 295 K (Fig. 2) and also signs of chemical reactivity through an appearance of 4520 $cm^{-1}$ broad peak above approximately 34 GPa, which increased in intensity and remained observable up to at least 110 GPa. The position of this peak suggests that it is related to vibron modes of unbound $H_2$ molecules observed by infrared spectroscopy [26]. Elemental S has not been clearly detected in any of the low-temperature runs. One of the most striking observations at 180 K is the formation of a long-range disordered state at 5-10 GPa when irradiated by X-rays at low temperatures (Fig. 4), which persists to 132 GPa (B, D, and E) even after warming up to 295 K. This radiation-induced low-temperature amorphization is another sign of the instability of $H_2S$ under pressure.

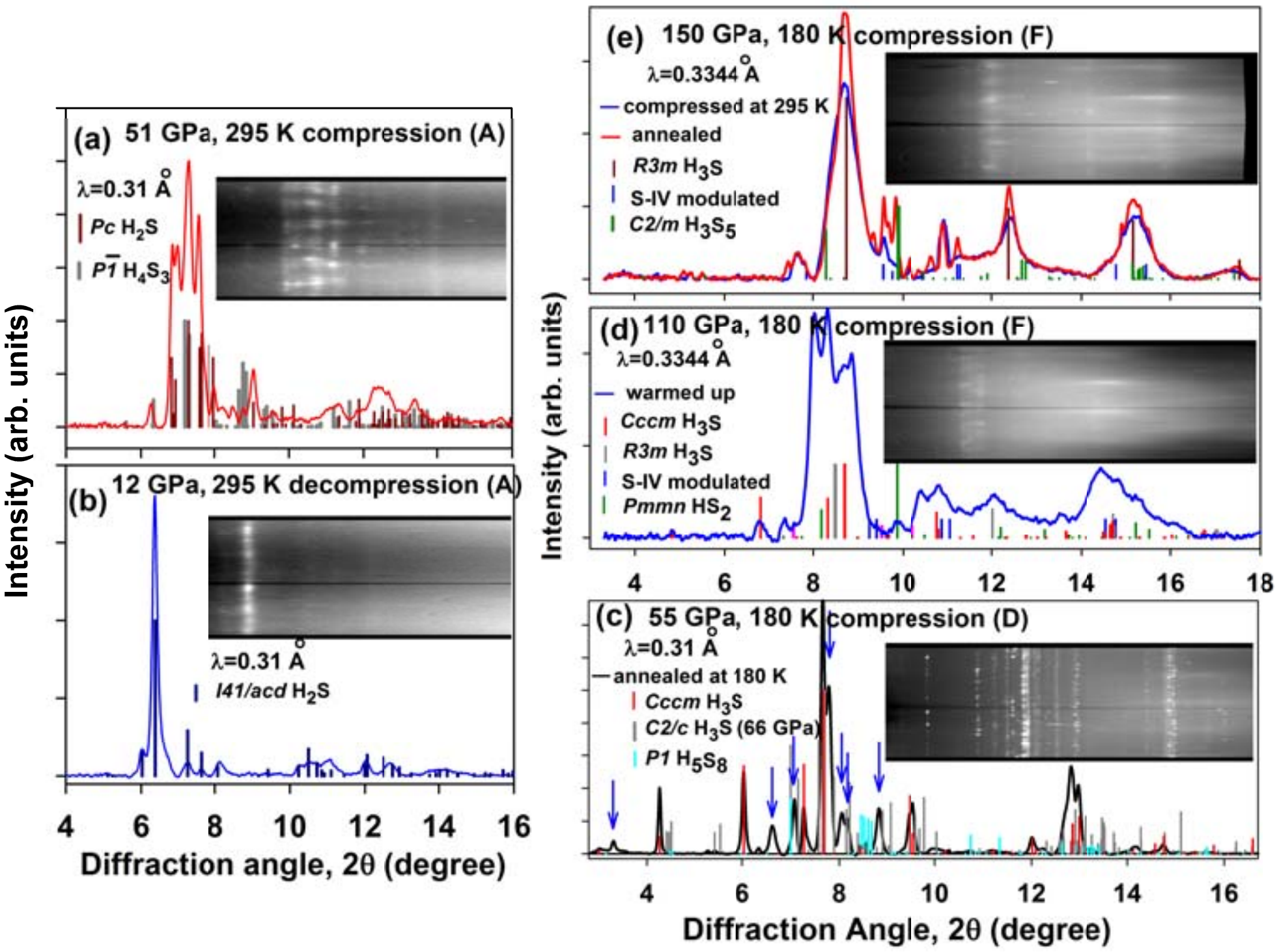

Fig. 3. XRD at 295 K of $H_2S$ samples compressed to high pressures at 295 (a, b) and 180 K (c, d, e). (a) at 51 GPa (experiment A); (b) decompressed to 12 GPa; (c) laser annealed at 180 K and then warmed up to 295 K at 55 GPa (D); (d) warmed up at 110 GPa (F); (e) subsequently compressed at 295 K (blue curve) and gently laser annealed (red curve). The lines are the measured XRD (background subtracted); vertical bars show the positions and intensities (arb. units) of Bragg peaks calculated using the inferred structures from this work and Refs. [3, 19, 23, 25]. The X-ray wavelengths used are specified in each panel. Vertical arrows in (c) show the positions of the Bragg peaks corresponding to a second phase, indexed as a monoclinic (Table 1 in SM [27]). Insets in all panels ((a) through (e)) show the raw diffraction images in rectangular coordinates (cake).

We annealed the amorphous samples at 55 and 110 GPa (D and E) by applying a gentle laser heating (<1000 K) at 180 K. The material completely re-crystallizes (Fig. 3(c)) revealing a rich XRD pattern with narrow and well defined Bragg peaks. The XRD image clearly shows that there are two sets of diffraction rings: spotty (larger crystallites) and quasi-continuous ones. The positions of spotty XRD rings correspond well to the theoretically predicted *Cccm* structure of $H_3S$ [3], which is closely related to the *I4/mcm* $(H_2S)_2H_2$ [28] observed below 30 GPa. Please note that the predicted in Ref. [19] *C2/c* $H_3S$ has been reported to have a slightly low enthalpy, which is supported in our calculations. However, *C2/c* $H_3S$ (which is expected to show distinct Bragg peaks) has not been detected in our XRD experiments (Fig. 3) that favor the formation of *Cccm* structure of $H_3S$ [3]. The refined structural parameters of *Cccm* $H_3S$ yield the density, which compares well with the extrapolated equation of state of $(H_2S)_2H_2$ [28] and results of theoretical calculations of Ref. [3] (Fig. 5). The second set of reflections corresponds to an unknown $H_xS$ phase. As it should be compositionally balanced with a hydrogen-rich *Cccm* $H_3S$, one can expect it to be a sulfur-rich compound with the density larger than that of pure $H_2S$ and smaller than that of pure S. We indexed the observed Bragg reflections (Table 1 in Supplemental Materials (SM) [27]) using this density constrain obtaining monoclinic solutions with 6 S atoms in the cell and densities clustering around 4.2(1) g/cm$^3$. Based on this determination, by comparing with the densities calculated theoretically for a number of stable $H_xS$ compounds (Fig. 5), we suggest the H content of x=0.5-1.0. At 110 GPa, the data can be indexed by an orthorhombic and monoclinic unit cells (Table 1 in SM [27]) again suggesting a second sulfur-rich H-S compound.

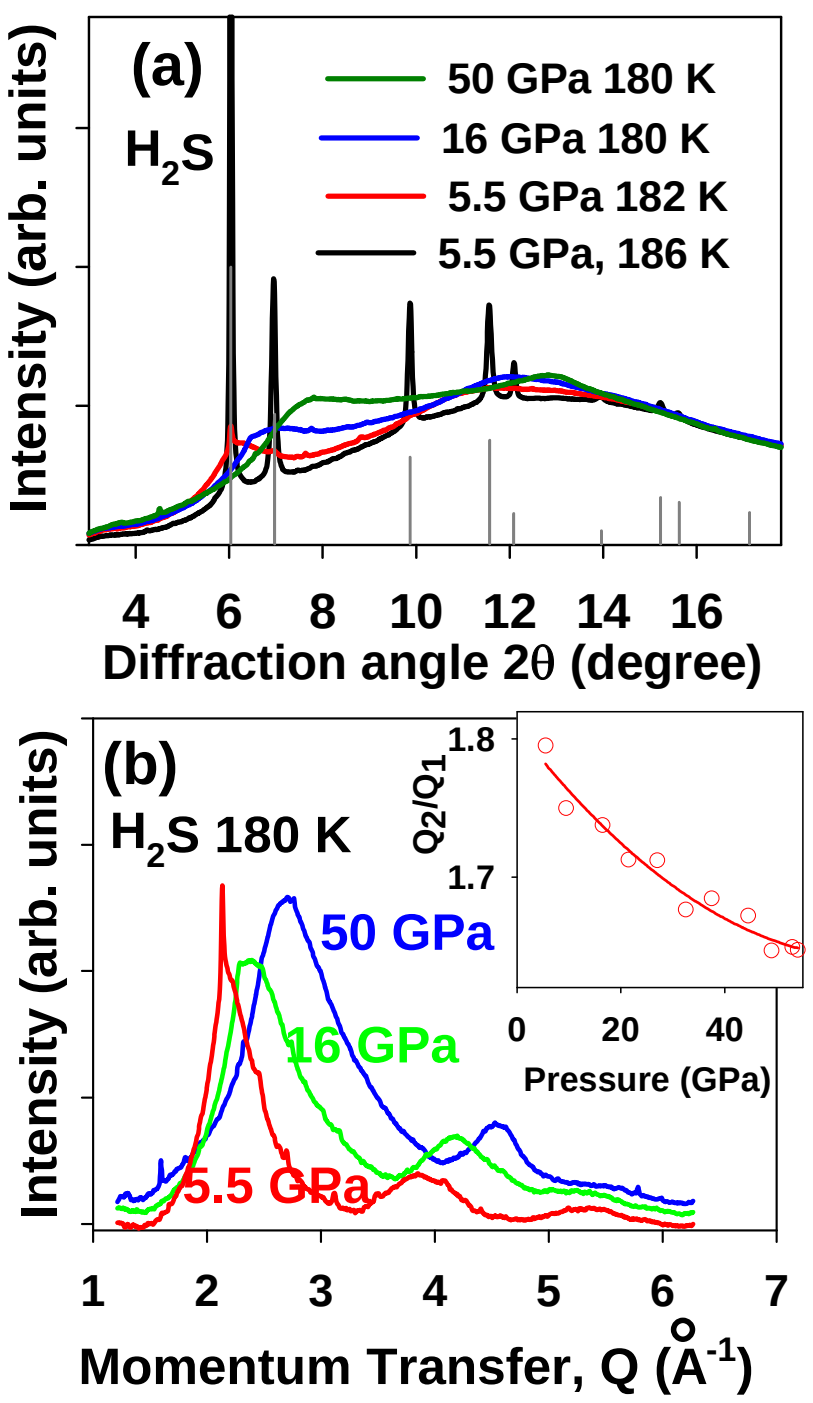

Fig. 4. X-ray induced amorphization of $H_2S$. (a) a series of XRD patterns of $H_2S$ taken on cooling down at 5.5 GPa to 180 K with a subsequent pressure increase to 55 GPa at 180 K (experiment D); the vertical bars are the calculated positions and intensities of the diffraction

peaks of a face centered cubic (fcc) $H_2S$-I (b) Broad diffraction peaks (background subtracted) of an amorphous state to 50 GPa plotted as a function of the momentum transfer $Q=4\pi\sin(\theta)/\lambda$, where $\lambda=0.31$ Å is the X-ray wavelength, $2\theta$ is the diffraction angle; an inset shows a pressure behavior of the ratio of the second to the first diffuse peak positions.

In experiments C, F, and G we studied XRD after pressurizing at 180 K and warming up (to avoid amorphization). The 55 GPa XRD pattern (C) was qualitatively similar to that obtained for the sample compressed at 295 K suggesting that predominantly $H_2S$-VI was present. Laser annealing resulted in a dramatic change of diffraction pattern; the observed Bragg peaks resemble those of the *Cccm* phase created after laser annealing of amorphous phase at 55 GPa (D). Experiments to 150 and 144 GPa (F and G) probed the experimental conditions in which superconducting properties have been reported [7, 8]. At 110 GPa after compressing at 180 K and warming up to 295 K (F), similarly to the case of the sample annealed at 110 GPa (E), we find several XRD peaks which match again *Cccm* $H_3S$ (Fig. 3(d)); the volume of the unit cell yields a density consistent with the data of experiment E (Fig. 5). The remaining peaks are broad, and the major peaks at 8.5 and 14.5 degrees are similar in position to those observed in an amorphous X-ray induced state (Fig. 4). The three major diffraction peaks match well those of *R3m* $H_3S$ structure [3], while the deduced density is also in a good agreement with theoretical calculations of this work and Ref. [3] (Fig. 5). Application of pressure at 295 K results in a continuous narrowing down of the major peaks and appearance of a forth broad peak at 17 degrees as well as appearance of several single-crystal spots, which become stronger after a laser annealing at 150 GPa. The peaks of *Cccm* $H_3S$ are reduced in intensity and cannot be seen at highest pressures.

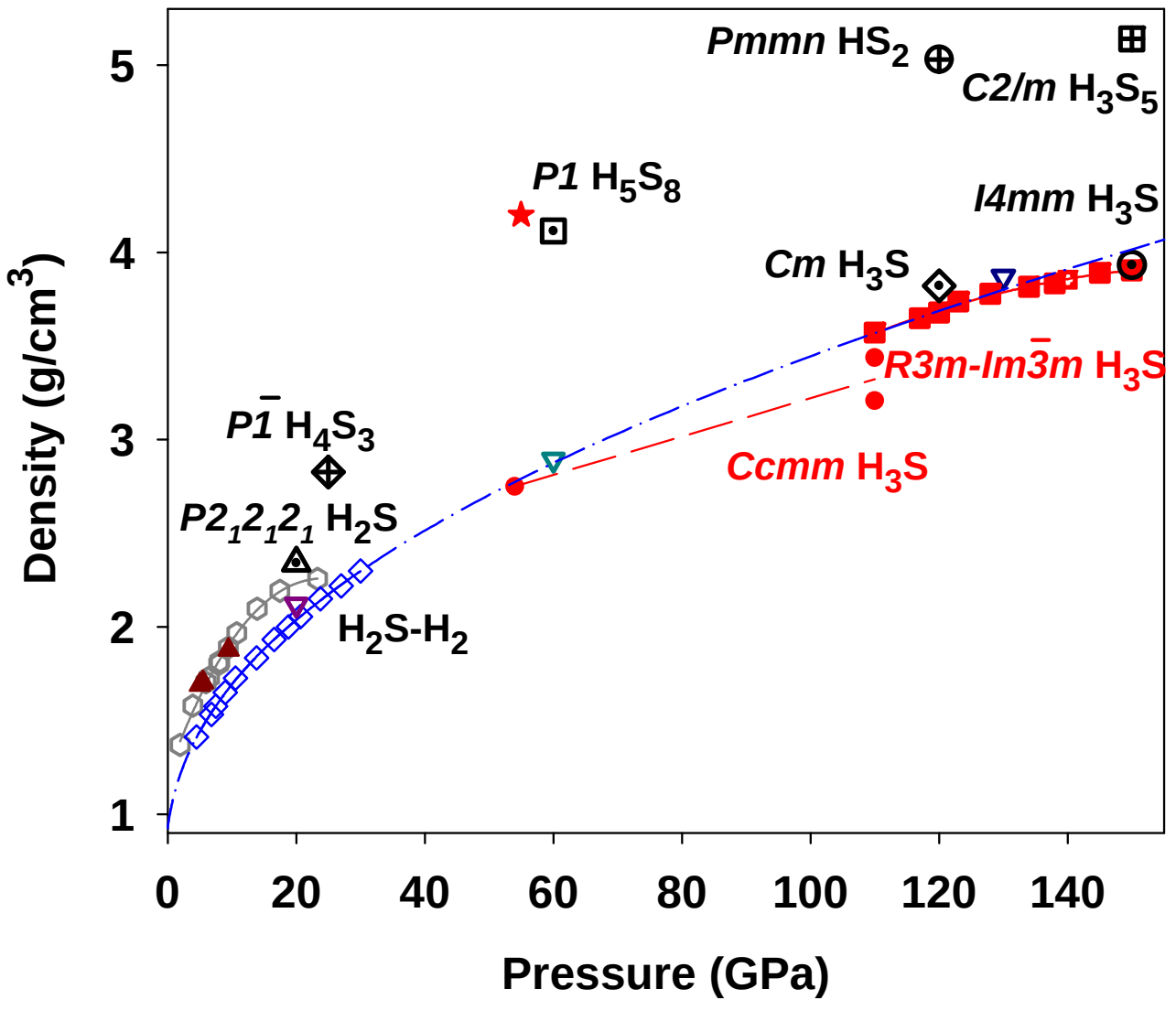

Fig. 5. The pressure–density plot for various $H_xS$ compounds. Open diamonds with a dot-dashed line extrapolated (using a Vinet fit to the existing data) to lower and higher

pressures correspond to $H_2S$-$H_2$ compounds [28]. Open hexagons (a solid line is a guide to the eye) are results for phases I, I', and IV [23], which agree well with our data (experiment A) obtained on unloading at 295 K (filled red triangles). Open triangles pointing down are the results of theoretical calculation for $H_3S$ in *P1*, *Cccm*, and *R3m* structures [3]. Red filled circles (a dashed line is a guide to the eye) correspond to *Cccm* structure refined in our experiments for samples compressed at low temperatures (D, E, F). Filled red squares correspond to *R3m* structure refined in our experiment after compressing to 110 GPa at 180 K, heating up at 110 GPa and compressing again at 295 K (F). Large dotted and crossed symbols correspond to various stable structures (compositions labeled) theoretically predicted in this work (Table 2). A filled star is the estimated density for a monoclinic structure suggested in this work using indexing of experimental Bragg reflections marked by arrows in Fig. 3(c) (Table 1 in SM [27]).

The laser annealing at 150 GPa narrowed down the major diffraction peaks, and a refinement of *R3m* (or *Im-3m,* which cannot be distinguished) $H_3S$ structure can be performed using the four major diffraction peaks (Fig. 3(e)). In addition, one can assign narrow diffraction peaks to a modulated elemental S structure [29]. A few peaks remain unassigned, specifically peaks at 7.7, 8.3, 9.8, and 10.9 degrees. Our theoretical calculations suggest a *C2/m* $H_3S_5$ (Table 2) to be a stable compound at these conditions; the presence of this material can account for two of the extra Bragg peaks, but others remain not well explained. In the experiment to 144 GPa (G), broad peaks of *R3m* $H_3S$ were observed after heating up to 295 K and no peaks of *Cccm* $H_3S$ were detected.

Unlike the results of previous works [3, 9, 19], our theoretical structure searches of the most stable thermodynamically structures suggest the symmetry breaking of *R3m* and *Im-3m* $H_3S$ down to distorted *Cm* and *I4mm* structures, respectively (Table 2). This should result in a small splitting of the major Bragg peaks (Fig. 3(e)), which our XRD measurements cannot determine due the broadness of the observed peaks.

## IV. Discussion

Our extensive experimental and theoretical sampling of $H_2S$ at high pressures and variable temperatures showed instability of $H_2S$ at pressures as low as 5 GPa (X-ray initiated) and starting as low as 27 GPa without X-ray irradiation. The low- and room-temperature reaction products are different suggesting that kinetics and experimental conditions (*e.g.*, possible hydrogen leak) and X-ray induced chemistry play an important role in the reaction pathway. The reaction products obtained during low- and room-temperature P-T excursions are of low crystallinity evidenced by broad and rather radially uniform diffraction rings; however, minor phases form single crystals. The laser annealing treatment results in the formation of well crystallized homogeneous powder products. This further supports the existence of high-pressure chemical boundary; traversing this boundary results in sample heterogeneity as $H_2S$ becomes unstable giving rise to compounds with various stoichiometries. RS and XRD observations below 51 GPa at 295 K suggest that these chemical transformations are reversible.

The chemical reactivity is superimposed by high-pressure physical transformations, which are manifested by progressing molecular instability causing loss of molecular identity through the proton exchange and then formation of new atomic associations and even polymerization. We have identified in this work an important intermediate atomic-molecular state, which is similar to the theoretically predicted *Cccm* $H_3S$ [3]. This phase is expected to form a 3D network structure, which is strongly hydrogen bonded with channels filled by weakly bound $H_2$ molecules. Our XRD patterns are remarkably similar to this prediction (Fig. 3(c)), while our Raman spectroscopy data (Fig. 2) rather suggests the presence of a strongly bound S network and very disordered hydrogen sublattice. Indeed our Raman measurements show a pronounced peak near 400 $cm^{-1}$ for the annealed sample containing a substantial amount of *Cccm* $H_3S$ (D) suggesting either formation of symmetric hydrogen bonds [30] or network of transient S-S bonds [31].

Our higher pressure experiments intended to determine the composition and structure of a superconducting phase revealed a crossover to denser extended structures. Due to a large kinetic barrier expected for the *Cccm* - *R3m* transition in $H_3S$ [3], the transitions is expected to be sluggish in a good accord with our observations, which showed the transition pressure range and phase coexistence at 295 K extended to several tens of GPa. Our measurements show a large densification (about 8%) at this transformation (Fig. 5) in a good agreement with the predicted 9.2%.

The presence of elemental sulfur at 150 GPa as one of secondary phases is definitive based on our XRD data (Fig. 3(e)). However, there are additional sulfur rich $H_xS$ phases, which we have detected but identifications of them remains difficult (cf. Ref. [12]). We also find that five structures predicted theoretically here, *P-1* $H_4S_3$, *P1* $H_5S_8$, *Pmmn* $HS_2$, and *C2/m* $H_3S_5$ at 51, 55, 110, and 150 GPa, respectively (Fig. 3, Table I) are broadly consistent with experiments. Using a different search technique, $H_3S$, $H_2S_3$, $H_3S_2$, and $HS_2$ structures were predicted [19], but disagree with our XRD patterns. Our prediction for $HS_2$ compound is different of that in Ref. [9] (albeit at higher pressures), which also finds a stable HS compounds above 300 GPa.

## V. Conclusions

Our experiments on $H_2S$ following different P-T paths (Fig. 1) show rich physical and chemical transformations detected with XRD and Raman spectroscopy combined with first principles theoretical structure searches. We found that chemical decomposition of $H_2S$ starts at approximately 30 GPa, moreover X-ray induced amorphization has been detected below 200 K at >10 GPa. The major reaction product below 120 GPa is found to be *Cccm* $H_3S$ [3]; other minor S-rich phases were difficult to identify. Although not performed simultaneously with detection of superconductivity, our experiments to highest pressures (F, G) were following P-T paths of Refs. [7, 8], and thus can be considered as a good probe of superconducting phase (Fig. 1). We found that at 140-150 GPa, *R3m* (*Im-3m*) $H_3S$ is indeed the major phase when following a low-temperature compression path thus confirming that *R3m* and/or *Im-3m* phases are responsible for high-$T_c$ superconductivity. However, our measurements clearly show the sluggishness of the transition to these phases and possible low-crystallinity of materials prepared by compression at low temperatures, which could affect the superconducting properties and can explain why annealing at 295 K can increase the $T_c$ [8]. The broadness of the diffraction peaks of the superconducting phase (Fig. 3(d, e)) could also be related to a possible symmetry breaking as determined by in our theoretical calculations (Table 2).

**V. Acknowledgement**

We thank Viktor Struzhkin, Maddury Somayazulu, and Zack Geballe for help in sample loading, many fruitful discussions, and comments on the manuscript. We acknowledge support from the ARO (No. W911NF-13-1-0231), DARPA (No. W31P4Q1210008), NSF (No. EAR-1520648 and 1531583), and NSFC (No. 21473211). X-ray diffraction experiments were performed at GeoSoilEnviroCARS (Sector 13), Advanced Photon Source (APS), Argonne National Laboratory and Petra III, DESY, Hamburg, Germany. GeoSoilEnviroCARS is supported by the National Science Foundation—Earth Sciences (No. EAR-1128799) and Department of Energy—Geosciences (No. DE-FG02-94ER14466). Use of the Advanced Photon Source was supported by the U. S. Department of Energy, Office of Science, Office of Basic Energy Sciences, under Contract No. DE-AC02-06CH11357. PETRA III at DESY is a member of the Helmholtz Association (HGF). Theoretical calculations were supported by the National Science Foundation (Nos. EAR-1114313 and DMR-1231586), DARPA (Grant Nos. W31P4Q1210008 and W31P4Q1310005), the Government (No. 14.A12.31.0003) and the Ministry of Education and Science of Russian Federation (Project No. 8512) for financial support, and Foreign Talents Introduction and Academic Exchange Program (No. B08040). Calculations were performed on XSEDE facilities and on the cluster of the Center for Functional Nanomaterials, Brookhaven National Laboratory, which is supported by the DOE-BES under Contract No. DE-AC02-98CH10086. The research leading to these results has received funding from the European Community's Seventh Framework Programme (No. FP7/2007-2013) under grant Agreement No. 312284. S.S.L. was partly supported by the Ministry of Education and Science of Russian Federation (Grant No. 14.B25.31.0032).

**Table 1. Experimental conditions of $H_2S$ experiments.**

| Experiment | Maximum Pressure | Pressure measurement technique | T path (K) | Probe | Laser annealing: starting temperature |
|---|---|---|---|---|---|
| A | 51 | Ruby, Raman of diamond | 295 | Raman (up), XRD (down) | None |
| B | 132 | XRD of Au | 190 | XRD | None |
| C | 55 | Raman of diamond | 180 | Raman | 295 |
| D | 55 | XRD of Au | 180 | XRD | 180 & 295 |
| E | 110 | XRD of Au | 180 | XRD | 180 & 295 |
| F | 150 | Raman of diamond | 180 & 295 | Raman to 110 GPa, XRD above 110 GPa | 295 at 150 GPa |
| G | 144 | Raman of diamond | 180 | Raman to 144 GPa XRD at 144 GPa | 295 |

**Table 2. Structural parameters of the high pressures $H_xS$ compounds theoretically predicted in this work.**

| Composition/ pressure | Space group | Lattice parameters (A, °) | Atom | Coordinates (fractional) | | |
|---|---|---|---|---|---|---|
| $H_2S$ (20 GPa) | *P2₁2₁2₁* (SG 19) | a=6.754<br>b=3.918<br>c=3.654 | H1<br>H2<br>S1 | 0.01782<br>0.01761<br>0.15798 | 0.42124<br>-0.42597<br>0.24933 | -0.19609<br>0.30542<br>0.00090 |
| $H_4S_3$ (25 GPa) | *P-1* (SG 2) | a=5.25400<br>b=5.15100<br>c=4.40800<br>α=99.3080<br>β=90.9840<br>γ=89.6370 | H1<br>H2<br>H3<br>H4<br>S1<br>S2<br>S3 | 0.02852<br>0.06166<br>0.26479<br>-0.45416<br>0.25296<br>0.25007<br>0.28163 | -0.20245<br>-0.35693<br>0.00161<br>0.36823<br>0.47722<br>0.15607<br>-0.15059 | -0.15400<br>0.32886<br>0.31188<br>-0.33827<br>0.19946<br>-0.40034<br>-0.13637 |
| $H_5S_8$ (60 GPa) | *P1* (SG 1) | a=4.30500<br>b=4.35600<br>c=6.16200<br>α=73.46600<br>β= 74.22100<br>γ= 92.86400 | H1<br>H2<br>H3<br>H4<br>H5<br>S1<br>S2<br>S3<br>S4<br>S5<br>S6<br>S7<br>S8 | 0.16163<br>0.47574<br>0.16598<br>-0.22060<br>0.49595<br>0.36226<br>-0.17866<br>-0.15969<br>0.35738<br>-0.16654<br>0.29610<br>0.29928<br>-0.17217 | 0.27713<br>0.03618<br>-0.17119<br>-0.25039<br>0.45654<br>-0.05860<br>0.40358<br>-0.06375<br>0.42879<br>0.41848<br>-0.11901<br>0.40212<br>-0.09852 | 0.38770<br>0.29369<br>-0.38874<br>-0.16293<br>0.07637<br>-0.27945<br>-0.29174<br>-0.02803<br>0.47848<br>0.21014<br>0.19657<br>-0.05669<br>0.45249 |
| $HS_2$ (120 GPa) | *Pmmn* (SG 59, origin choice 2) | a=3.14700<br>b=2.91500<br>c=4.68800 | H1<br>S1<br>S2 | 0.25000<br>0.25000<br>0.25000 | 0.25000<br>0.75000<br>0.25000 | 0.25804<br>-0.00265<br>-0.45788 |
| $H_3S$ (120 GPa) | *Cm* (SG 8) | a=3.14856<br>b=4.40409<br>c=2.70291<br>β=125.53483 | H1<br>H2<br>S1 | -0.19742<br>0.13876<br>-0.42124 | 0.23764<br>0.00000<br>0.00000 | -0.36944<br>0.15491<br>0.14307 |
| $H_3S_5$ (150 GPa) | *C2/m* (SG 12) | a=7.61911<br>b=3.03000<br>c=7.11444<br>β=140.01883 | H1<br>H2<br>S1<br>S2<br>S3 | 0.34694<br>0.00000<br>-0.20492<br>-0.40306<br>0.00000 | 0.00000<br>0.00000<br>0.00000<br>0.00000<br>0.00000 | -0.09445<br>0.00000<br>-0.32241<br>-0.13707<br>0.50000 |
| $H_3S$ (150 GPa) | *I4mm* (SG 107) | a=b=3.08477<br>c=3.11415 | H1<br>H2<br>S1 | 0.0<br>0.0<br>0.0 | 0.5<br>0.0<br>0.0 | 0.09379<br>0.16267<br>-0.38691 |

**References and Notes:**

1. N. W. Ashcroft, Phys. Rev. Lett. **92**, 187002 (2004).
2. E. Zurek, R. Hoffmann, N. W. Ashcroft, A. R. Oganov, and A. O. Lyakhov, Proc. Natl. Acad. Sci. U.S.A. **106**, 17640 (2009).
3. D. Duan, Y. Liu, F. Tian, D. Li, X. Huang, Z. Zhao, H. Yu, B. Liu, W. Tian, and T. Cui, Sci. Rep. **4**, 6968 (2014).
4. C. Pépin, P. Loubeyre, F. Occelli, and P. Dumas, Proc. Natl. Acad. Sci. U.S.A. **112**, 7673 (2015).
5. V. V. Struzhkin, D.-Y. Kim, E. Stavrou, T. Muramatsu, H.-K Mao, C. J. Pickard, R. J. Needs, V. B. Prakapenka, and A. F. Goncharov, arXiv:1412.1542, Nat. Comm, in press.
6. M. I. Eremets, I. A. Trojan, S. A. Medvedev, J. S. Tse, and Y. Yao, Science **319**, 1506 (2008).
7. A. P. Drozdov, M. I. Eremets and I. A. Troyan, arXiv:1412.0460.
8. A. P. Drozdov, M. I. Eremets, I. A. Troyan, V. Ksenofontov and S. I. Shylin, Nature **525**, 73 (2015).
9. I. Errea, M. Calandra, C. J. Pickard, J. Nelson, R. J. Needs, Y. Li, H. Liu, Y. Zhang, Y. Ma, and F. Mauri, Phys. Rev. Lett. **114**, 157004 (2015).
10. N. Bernstein, C. S. Hellberg, M. D. Johannes, I. I. Mazin, and M. J. Mehl, Phys. Rev. B **91**, 060511 (2015).
11. J. A. Flores-Livas, A. Sanna and E. K. U. Gross, arXiv:1501.06336.
12. D. Duan, X. Huang, F. Tian, D. Li, H. Yu, Y. Liu, Y. Ma, B. Liu, and T. Cui, Phys. Rev. B **91**, 180502 (2015).
13. R. Akashi, M. Kawamura, S. Tsuneyuki, Y. Nomura, and R. Arita, Phys. Rev. B **91**, 224513 (2015).
14. I. Errea, M. Calandra, C. J. Pickard, J. R. Nelson, R. J. Needs, Y. Li, H. Liu, Y. Zhang, Y. Ma & F. Mauri,. Nature, **532**, 81 (2016).
15. M. Sakashita, H. Fujihisa, H. Yamawaki, and K. Aoki, J. Phys. Chem. A **104**, 8838 (2000).
16. M. Sakashita, H. Yamawaki, H. Fujihisa, K. Aoki, S. Sasaki, and H. Shimizu, Phys. Rev. Lett. **79**, 1082 (1997).
17. H. Fujihisa, H. Yamawaki, M. Sakashita, A. Nakayama, T. Yamada, and K. Aoki, Phys. Rev. B **69**, 214102 (2004).
18. H. Shimizu, T. Ushida, S. Sasaki, M. Sakashita, H. Yamawaki, and K. Aoki, Phys. Rev. B **55**, 5538 (1997).
19. Y. Li, L. Wang, H. Liu, Y. Zhang, J. Hao, C. J. Pickard, J. R. Nelson, R. J. Needs, W. Li, Y. Huang, I. Errea, M. Calandra, F. Mauri and Y. Ma, Phys. Rev. B **93**, 020103 (2016).
20. A. R. Oganov,Y. M. Ma, A. O. Lyakhov, M.Valle, and C. Gatti, Rev. Mineral. Geochem. **71**, 271 (2010).
21. J. P. Perdew, K. Burke, and M. Ernzerhof, Phys. Rev. Lett. **77**, 3865 (1996).
22. G. Kresse and J. Furthmuller, Comput. Mater. Sci. **6**, 15 (1996).
23. H. Fujihisa, H. Yamawaki, M. Sakashita, K. Aoki, S. Sasaki, and H. Shimizu,

Phys. Rev. B **57**, 2651 (1998).
24. S. Endo, A. Honda, S. Sasaki, H. Shimizu, O. Shimomura and T. Kikegawa, Phys. Rev. B **54**, R717 (1996).
25. Y. Li, J. Hao, H. Liu, Y. Li, and Y. Ma, J. Chem. Phys. **140**, 174712 (2014).
26. M. Hanfland, R. J. Hemley, H.-k. Mao, and G. P. Williams, Phys. Rev. Lett. **69**, 1129 (1992).
27. See Supplemental Material at http://link.aps.org/supplemental/10.1103/PhysRevB.xxx for indexing of X-ray diffraction patterns in experiments D and E.
28. T. A. Strobel, P. Ganesh, M. Somayazulu, P. R. C. Kent, and R. J. Hemley, Phys. Rev. Lett. **107**, 255503 (2011).
29. O. Degtyareva, E. Gregoryanz, M. Somayazulu, H.-k. Mao, and R. J. Hemley, Phys. Rev. B **71**, 214104 (2005).
30. A. F. Goncharov, V. V. Struzhkin, H.-k. Mao, and R. J. Hemley, Phys. Rev. Lett. **83**, 1998 (1999).
31. R. Rousseau, M. Boero, M. Bernasconi, M. Parrinello, and K. Terakura, Phys. Rev. Lett. **85**, 1254 (2000).